\documentclass{article}
\usepackage{cite}
\usepackage{graphicx}
\usepackage{dcolumn}

\begin{document}

\title{Simple model for Crystal Field Theory}
\author{Francisco M. Fern\'{a}ndez\\
INIFTA (UNLP, CCT La Plata-CONICET), Divisi\'on Qu\'imica
Te\'orica\\ Blvd. 113 S/N,  Sucursal 4, Casilla de Correo 16, 1900
La Plata, Argentina}

\maketitle

\begin{abstract}
We investigate a simple model for the prediction of the splitting of the $3d$
orbitals of a metal ion in the environment of ligands. The electrons are
considered to be independent and their interaction with the ligands is
represented by the Dirac delta function in three dimensions. We discuss
several cases where the model is successful and a few ones where it is not.
\end{abstract}

\section{Introduction}

The crystal field theory (CFT) developed by Bethe\cite{B29} long
ago was utilized by physicists to explain magnetic properties and
absorption spectra of transition metals and other compounds. This
theory is commonly discussed in specialized
textbooks\cite{B62,B93} as well as in standard books on inorganic
chemistry\cite{CW72}.

CFT treats the interaction between the metal ion and the ligands as a purely
electrostatic problem in which the ligand atoms or molecules are considered
to be mere point charges (or point dipoles). In spite of this
oversimplification CFT provides a very simple and easy way of treating
numerically many aspects of the electronic structure of complexes.

The purpose of this paper is to investigate an even simpler CFT model were
the electrostatic repulsion between the $3d$ electrons (assumed to be
noninteracting) and the ligands is represented by a Dirac delta function in
three dimensions centered at the ligand positions. Although this extremely
short-range interaction may appear to be rather too unrealistic at first
sight some of the results are surprisingly reasonable.

In section~\ref{sec:model} we develop the model that we later apply to some
examples in section~\ref{sec:examples}. We consider cases where the model is
successful and others in which it fails. Finally we summarize the main
results and draw conclusions in section~\ref{sec:conclusions}.

\section{The model}

\label{sec:model}

The model takes into account only $3d$ electrons that are
considered to be independent. It is further assumed that in the
absence of ligands each electron can be described by an effective
hydrogen-like Hamiltonian
\begin{equation}
H_{0}=-\frac{\hbar ^{2}}{2m}\nabla ^{2}-\frac{Ze^{2}}{4\pi \epsilon _{0}r},
\label{eq:H0}
\end{equation}
where $Z$ is an effective core charge. The effect of $M$ negative ligands
around the central metal ion is taken into account by means of a
perturbation Hamiltonian of the form
\begin{equation}
H^{\prime }=V_{0}\sum_{k=1}^{M}\delta (\mathbf{r}-\mathbf{r}_{k}),
\label{eq:H'}
\end{equation}
where $V_{0}>0$ is the strength of the electrostatic interaction
between a $d$ electron and the ligands and $\delta
(\mathbf{r}-\mathbf{r}_{k})$ is the well known Dirac delta
function in three dimensions. In the absence of ligands the $3d$
hydrogen-like orbitals
\begin{equation}
\left\{ \varphi _{j},\;j=1,2,\ldots 5\right\} =\left\{ \varphi
_{d_{z^{2}}},\varphi _{d_{x^{2}-y^{2}}},\varphi _{d_{xz}},\varphi
_{d_{yz}},\varphi _{d_{xy}}\right\} ,  \label{eq:orbitales}
\end{equation}
are degenerate.

In order to calculate the splitting of the 5-fold degenerate $3d$ level we
resort to perturbation theory. The perturbation corrections of first order
to the orbital energies are given by the roots $\epsilon _{n}^{\prime }$, $%
n=1,2,3,4,5$ of the secular determinant
\begin{equation}
\left| \mathbf{H}^{\prime }-\epsilon ^{\prime }\mathbf{I}\right| =0,
\label{eq:secular_det}
\end{equation}
where the elements of the perturbation matrix $\mathbf{H}^{\prime }$ are
\begin{equation}
H_{ij}^{\prime }=\left\langle \varphi _{i}\right| H^{\prime }\left| \varphi
_{j}\right\rangle =V_{0}\sum_{k=1}^{M}\varphi _{i}(\mathbf{r}_{k})\varphi
_{j}(\mathbf{r}_{k}).  \label{eq:H'ij}
\end{equation}
Note that the calculation reduces to obtaining the values of each $3d$
orbital at the locations of the ligands.

The calculation is greatly simplified if we write the $3d$ real orbitals as%
\cite{EWK44}
\begin{eqnarray}
\varphi _{j}(x,y,z) &=&f(\rho )u_{j}(x,y,z)  \nonumber \\
f(\rho ) &=&\frac{\sqrt{2}}{81a_{0}^{2}\sqrt{\pi }}\left( \frac{Z}{a_{0}}%
\right) ^{3/2}e^{-\rho /3},\;\rho =\frac{Zr}{a_{0}},  \label{eq:orbitales_2}
\end{eqnarray}
where $a_{0}$ is the Bohr radius and the form factors $u_{j}(x,y,z)$ read
\begin{equation}
\begin{array}{|l|l|l|l|l|l|}
\mathrm{Orbital} & d_{z^{2}} & d_{x^{2}-y^{2}} & d_{xz} & d_{yz} & d_{xy} \\
\hline
u & \frac{1}{\sqrt{12}}\left( 2z^{2}-x^{2}-y^{2}\right)  & \frac{1}{2}\left(
x^{2}-y^{2}\right)  & xz & yz & xy
\end{array}
.  \label{eq:orbitales_3}
\end{equation}
In this way, the matrix elements (\ref{eq:H'ij}) can be written in the
simpler form
\begin{eqnarray}
H_{ij}^{\prime } &=&\frac{\lambda }{L^{4}}\sum_{k=1}^{M}u_{i}\left(
x_{k},y_{k},z_{k}\right) u_{j}\left( x_{k},y_{k},z_{k}\right) ,  \nonumber \\
\lambda  &=&V_{0}f(ZL/a_{0})^{2}L^{4},  \label{eq:H'ij_2}
\end{eqnarray}
where $L$ is the distance between each one of the ligands and the metal
nucleus. For the time being we assume that all such distances are equal in
order to simplify the presentation.

It follows from the Hellmann-Feynman theorem
\begin{equation}
\frac{d\epsilon }{dV_{0}}=V_{0}^{-1}\left\langle \psi \right| H^{\prime
}\left| \psi \right\rangle =\sum_{k=1}^{M}\psi \left( \mathbf{r}_{k}\right)
^{2}\geq 0,
\end{equation}
that $\epsilon \left( V_{0}\right) \geq \epsilon (0)$. Therefore, the
matrices constructed in the way just indicated do not satisfy the so called
\textit{center of gravity} rule\cite{B93,CW72}. Although this rule is
irrelevant for our purposes we can build matrices that satisfy it within
present simple model. For example, the modified perturbation matrix
\begin{equation}
\mathbf{H}_{cg}^{\prime }=\mathbf{H}^{\prime }-\frac{1}{5}tr\left( \mathbf{H}%
^{\prime }\right) \mathbf{I,}
\end{equation}
where $tr$ denotes the trace of a matrix and $\mathbf{I}$ is the $5\times 5$
identity matrix, already satisfies the center of gravity rule.

\section{Examples}

\label{sec:examples}

In what follows we test the predictive power of the model outlined
in the preceding section on several complexes of various
geometries.

\subsection{Tetrahedral complex}

In the case of four ligands at the vertexes of a tetrahedron we
have
\begin{equation}
\mathbf{r}_{s}=\frac{L}{\sqrt{3}}\left(
\begin{array}{ccc}
1 & 1 & 1 \\
-1 & -1 & 1 \\
-1 & 1 & -1 \\
1 & -1 & -1
\end{array}
\right) ,  \label{eq:r_tetra}
\end{equation}
where each row of the structure matrix $\mathbf{r}_{s}$ is one of the ligand
position vectors $\mathbf{r}_{k}$ with modulus $\left| \mathbf{r}_{k}\right|
=L$. A straightforward calculation leads to the following diagonal
perturbation matrix:
\begin{equation}
\mathbf{H}^{\prime }=\lambda \left(
\begin{array}{ccccc}
0 & 0 & 0 & 0 & 0 \\
0 & 0 & 0 & 0 & 0 \\
0 & 0 & 4/9 & 0 & 0 \\
0 & 0 & 0 & 4/9 & 0 \\
0 & 0 & 0 & 0 & 4/9
\end{array}
\right) .  \label{eq:H'_tetra}
\end{equation}
We appreciate that the model predicts the widely accepted splitting of the
energy levels $\epsilon \left( d_{z^{2}}\right) =\epsilon \left(
d_{x^{2}-y^{2}}\right) <\epsilon \left( d_{xz}\right) =\epsilon \left(
d_{yz}\right) =\epsilon \left( d_{xy}\right) $\cite{B62,B93,CW72} with the
energy interval $\epsilon (T_{2})-\epsilon (E)=\Delta _{T}=4\lambda /9$.

\subsection{Octahedral complex}

The octahedral structure matrix
\begin{equation}
\mathbf{r}_{s}=L\left(
\begin{array}{ccc}
1 & 0 & 0 \\
0 & -1 & 0 \\
-1 & 0 & 0 \\
0 & 1 & 0 \\
0 & 0 & 1 \\
0 & 0 & -1
\end{array}
\right) ,  \label{eq:r_octa}
\end{equation}
describes four ligands located on the $xy$ plane at the vertexes
of a square and two more at opposite sides on the $z$ axis. It
leads to the diagonal perturbation matrix
\begin{equation}
\mathbf{H}^{\prime }=\lambda \left(
\begin{array}{lllll}
1 & 0 & 0 & 0 & 0 \\
0 & 1 & 0 & 0 & 0 \\
0 & 0 & 0 & 0 & 0 \\
0 & 0 & 0 & 0 & 0 \\
0 & 0 & 0 & 0 & 0
\end{array}
\right) ,  \label{eq:H'_octa}
\end{equation}
that exhibits the well known splitting of the energy levels
$\epsilon \left( d_{xz}\right) =\epsilon \left( d_{yz}\right)
=\epsilon \left( d_{xy}\right) <\epsilon \left( d_{z^{2}}\right)
=\epsilon \left( d_{x^{2}-y^{2}}\right) $. In this case the energy
interval is $\epsilon (T_{2g})-\epsilon (E_{g})=\Delta
_{O}=\lambda $. The most interesting feature of present simple
model is that it correctly predicts the ratio between the
tetrahedral and octahedral splittings $\Delta
_{T}=\frac{4}{9}\Delta _{O}$ that is commonly obtained by more
elaborate methods\cite{B62}. This ratio is valid provided that all
the model parameters in $\lambda $ ($Z$, $L$, $V_{0}$) are exactly
the same for both complexes.

\subsection{Cubic complex}

From the structure matrix that represents eight ligands on the vertexes of a
cube
\begin{equation}
\mathbf{r}_{s}=\frac{L}{\sqrt{3}}\left(
\begin{array}{ccc}
1 & 1 & 1 \\
-1 & 1 & 1 \\
-1 & -1 & 1 \\
1 & -1 & 1 \\
1 & -1 & -1 \\
1 & 1 & -1 \\
-1 & 1 & -1 \\
-1 & -1 & -1
\end{array}
\right) ,  \label{eq:r_cubic}
\end{equation}
we easily obtain the diagonal perturbation matrix
\begin{equation}
\mathbf{H}^{\prime }=\lambda \left(
\begin{array}{ccccc}
0 & 0 & 0 & 0 & 0 \\
0 & 0 & 0 & 0 & 0 \\
0 & 0 & 8/9 & 0 & 0 \\
0 & 0 & 0 & 8/9 & 0 \\
0 & 0 & 0 & 0 & 8/9
\end{array}
\right) ,  \label{eq:H'_cubic}
\end{equation}
that predicts the well known splitting $\epsilon \left( d_{z^{2}}\right)
=\epsilon \left( d_{x^{2}-y^{2}}\right) <\epsilon \left( d_{xz}\right)
=\epsilon \left( d_{yz}\right) =\epsilon \left( d_{xy}\right) $ and the
energy difference  $\epsilon (T_{2g})-\epsilon (E_{g})=\Delta _{C}=\frac{8}{9%
}\lambda $. Besides, the model predicts the correct ratio between
the magnitudes of the cubic and octahedral splittings $\Delta
_{C}=\frac{8}{9}\Delta _{O}$ obtained by means of more elaborate
methods\cite{B62}. Note that the ratios $\frac{9}{4}\Delta
_{T}=\Delta _{O}=\frac{9}{8}\Delta _{C}$ are mentioned in most
textbooks on the field\cite{B62,B93,CW72} but explicitly
calculated in a few of them\cite{B62}. Undoubtedly, present
calculation is far simpler.

\subsection{Square antiprismatic complex}

If we rotate the four ligands on one of the cube faces perpendicular to the $%
z$ axis by an angle $\pi /4$ about that axis we obtain the structure matrix
for a square antiprismatic complex
\begin{equation}
\mathbf{r}_{s}=\frac{L}{\sqrt{3}}\left(
\begin{array}{ccc}
0 & \sqrt{2} & 1 \\
-\sqrt{2} & 0 & 1 \\
0 & -\sqrt{2} & 1 \\
\sqrt{2} & 0 & 1 \\
1 & -1 & -1 \\
1 & 1 & -1 \\
-1 & 1 & -1 \\
-1 & -1 & -1
\end{array}
\right) .  \label{eq:sq-anti}
\end{equation}
A straightforward calculation yields
\begin{equation}
\mathbf{H}^{\prime }=\lambda \left(
\begin{array}{ccccc}
0 & 0 & 0 & 0 & 0 \\
0 & 4/9 & 0 & 0 & 0 \\
0 & 0 & 8/9 & 0 & 0 \\
0 & 0 & 0 & 8/9 & 0 \\
0 & 0 & 0 & 0 & 4/9
\end{array}
\right) ,  \label{eq:H'_sq-anti}
\end{equation}
that predicts the correct splitting $\epsilon \left( d_{z^{2}}\right)
<\epsilon \left( d_{x^{2}-y^{2}}\right) =\epsilon \left( d_{xy}\right)
<\epsilon \left( d_{xz}\right) =\epsilon \left( d_{yz}\right) $.

\subsection{Trigonal bipyramidal complex}

The structure matrix describing three ligands on the vertexes of an
equilateral triangle on the $xy$ plane and two more on the $z$ axis is given
by
\begin{equation}
\mathbf{r}_{s}=L\left(
\begin{array}{ccc}
1 & 0 & 0 \\
-1/2 & \sqrt{3}/2 & 0 \\
-1/2 & -\sqrt{3}/2 & 0 \\
0 & 0 & 1 \\
0 & 0 & -1
\end{array}
\right) .  \label{eq:r_trig}
\end{equation}
From it we obtain the diagonal perturbation matrix
\begin{equation}
\mathbf{H}^{\prime }=\lambda \left(
\begin{array}{ccccc}
11/12 & 0 & 0 & 0 & 0 \\
0 & 3/8 & 0 & 0 & 0 \\
0 & 0 & 0 & 0 & 0 \\
0 & 0 & 0 & 0 & 0 \\
0 & 0 & 0 & 0 & 3/8
\end{array}
\right) ,  \label{eq:H'_trig}
\end{equation}
that predicts the well known orbital splitting $\epsilon \left(
d_{xz}\right) =\epsilon \left( d_{yz}\right) <\epsilon \left(
d_{x^{2}-y^{2}}\right) =\epsilon \left( d_{xy}\right) <\epsilon \left(
d_{z^{2}}\right) $.

\subsection{Pentagonal bipyramidal complex}

If 5 ligands are located at the vertexes of a pentagon on the $xy$ plane and
two more on the $z$ axis the resulting structure matrix
\begin{equation}
\mathbf{r}_{s}=L\left(
\begin{array}{lll}
1 & 0 & 0 \\
\frac{\sqrt{5}-1}{4} & \sqrt{\frac{\sqrt{5}+5}{8}} & 0 \\
-\frac{\sqrt{5}-1}{4} & \sqrt{\frac{\sqrt{5}-5}{8}} & 0 \\
-\frac{\sqrt{5}+1}{4} & -\sqrt{\frac{\sqrt{5}-5}{8}} & 0 \\
\frac{\sqrt{5}-1}{4} & -\sqrt{\frac{\sqrt{5}+5}{8}} & 0 \\
0 & 0 & 1 \\
0 & 0 & -1
\end{array}
\right) ,  \label{eq:r_penta}
\end{equation}

leads to the diagonal perturbation one
\begin{equation}
\mathbf{H}^{\prime }=\lambda \left(
\begin{array}{ccccc}
13/12 & 0 & 0 & 0 & 0 \\
0 & 5/8 & 0 & 0 & 0 \\
0 & 0 & 0 & 0 & 0 \\
0 & 0 & 0 & 0 & 0 \\
0 & 0 & 0 & 0 & 5/8
\end{array}
\right) ,  \label{eq:H'_penta}
\end{equation}
that predicts the correct orbital splitting $\epsilon \left( d_{xz}\right)
=\epsilon \left( d_{yz}\right) <\epsilon \left( d_{x^{2}-y^{2}}\right)
=\epsilon \left( d_{xy}\right) <\epsilon \left( d_{z^{2}}\right) $.

\subsection{Square complex}

If there are four ligands at the vertexes of a square on the $xy$ plane the
structure matrix reads

\begin{equation}
\mathbf{r}_{s}=L\left(
\begin{array}{ccc}
1 & 0 & 0 \\
0 & -1 & 0 \\
-1 & 0 & 0 \\
0 & 1 & 0
\end{array}
\right) .  \label{eq:r_octdef}
\end{equation}
The symmetry elements of this complex are those of the point group
$D_{4h}$ and its character table predicts that the degeneracy of
the $d$ orbitals
should be $\left\{ d_{z^{2}}\right\} $, $\left\{ d_{x^{2}-y^{2}}\right\} $, $%
\left\{ d_{xy}\right\} $, $\left\{ d_{xz},d_{yz}\right\} $\cite{C90}.
Present oversimplified model leads to a somewhat greater degeneracy as shown
by the diagonal perturbation matrix
\begin{equation}
\mathbf{H}^{\prime }=\lambda \left(
\begin{array}{ccccc}
1/3 & 0 & 0 & 0 & 0 \\
0 & 1 & 0 & 0 & 0 \\
0 & 0 & 0 & 0 & 0 \\
0 & 0 & 0 & 0 & 0 \\
0 & 0 & 0 & 0 & 0
\end{array}
\right) .  \label{eq:H'_cuad}
\end{equation}
Note that the $d_{xy}$ orbital remains degenerate with $\left\{
d_{xz},d_{yz}\right\} $ in spite of the fact that the former and
latter are bases for the $B_{2g}$ and $E_{g}$ irreducible
representations, respectively\cite{C90}. Present model also fails
if we distort the octahedral complex along the $z$ axis leading to
either $D_{4h}$ or $C_{4v}$ symmetry\cite{C90}. The reason is that
the three orbitals $d_{xy}$, $d_{xz}$ and $d_{yz}$ vanish at all
ligand locations and are not affected by the distortion. We
conclude that the Dirac delta interaction appears to be unsuitable
for the description of these complexes

\section{Conclusions}

\label{sec:conclusions}

The model developed in this paper for the splitting of the $3d$
orbitals of a metal ion in a ligand environment is extremely
simple and easy to apply because it only requires the calculation
of the values of the $d$ orbitals at the ligand positions in
space. It provides the well known order of energy levels in most
of the cases and even yields the relative magnitudes of the
splittings. Although it fails to predict the splitting of the
orbitals in distorted octahedral complexes we think that it is an
interesting model for the qualitative (and even semi quantitative)
discussion of such molecules, particularly because the standard
approach is considerably more difficult to apply\cite{B62}.

\end{document}